
%
%
%
%
\newif\ifproofmode			
\proofmodefalse				

\newif\ifforwardreference		
\forwardreferencetrue			

\newif\ifeqchapternumbers		
\eqchapternumbersfalse			

\newif\ifsectionnumbers			
\sectionnumberstrue			

\newif\ifeqsectionnumbers		
\eqsectionnumbersfalse			

\newif\ifchaptersectionnumbers     	
\chaptersectionnumberstrue		

\newif\ifcontinuoussectionnumbers	
\continuoussectionnumbersfalse	

\newif\ifcontinuousnumbers		
\continuousnumbersfalse 		

\newif\iffigurechapternumbers		
\figurechapternumbersfalse		

\newif\ifcontinuousfigurenumbers	
\continuousfigurenumbersfalse		

\newif\ifcontinuousreferencenumbers     
\continuousreferencenumberstrue         

\newif\ifparenequations			
\parenequationstrue			

\newif\ifcrossreference			
\crossreferencefalse			

\newif\ifstillreading			

\font\eqsixrm=cmr6			
\def\marginstyle{\eqsixrm}		

\newtoks\chapletter			
\newcount\chapno			
\newcount\sectno			
\newcount\eqlabelno			
\newcount\figureno			
\newcount\referenceno			
\newcount\minutes			
\newcount\hours				

\newread\labelfile			
\newwrite\labelfileout			
\newwrite\allcrossfile			

\chapno=0
\sectno=0
\eqlabelno=0
\figureno=0


\def\chapternumberstrue{\eqchapternumberstrue}

%
\def\initialeqmacro{
    \ifproofmode
        \headline{\tenrm \today\ --\ \timeofday\hfill
                         \jobname\ --- draft\hfill\folio}
        \hoffset=-1cm
        \ifcrossreference
            \immediate\openout\allcrossfile=zallcrossreferfile
        \fi
    \else
        \crossreferencefalse
    \fi
    \ifforwardreference
        \openin\labelfile=zlabelfile
        \ifeof\labelfile
        \else
            \stillreadingtrue
            \loop
                \read\labelfile to \nextline
                \ifeof\labelfile
                    \stillreadingfalse
                \else
                    \nextline
                \fi
                \ifstillreading
            \repeat
        \fi
        \immediate\openout\labelfileout=zlabelfile
    \fi}


{\catcode`\^^I=9
\catcode`\ =9
\catcode`\^^M=9
\endlinechar=-1
\globaldefs=1


%
\def\chapfolio{			
    \ifnum \chapno>0 \relax
        \the\chapno
    \else
        \the\chapletter
    \fi}

%
\def\bumpchapno{
    \ifnum \chapno>-1 \relax
        \global \advance \chapno by 1
    \else
        \global \advance \chapno by -1 \setletter\chapno
    \fi
    \ifcontinuousnumbers
    \else
        \global\eqlabelno=0
    \fi
    \ifcontinuousfigurenumbers
    \else
        \global\figureno=0
    \fi
    \ifcontinuousreferencenumbers
    \else
        \global\referenceno=0
    \fi
    \sectno=0}

\def\bumpsectno{
    \global\advance\sectno by 1 \relax
    \ifeqsectionnumbers
        \ifcontinuoussectionnumbers
        \else
            \global\eqlabelno=0
        \fi
    \fi}

%
\def\setletter#1{\ifcase-#1 {}  \or\global\chapletter={A}
  \or\global\chapletter={B} \or\global\chapletter={C} \or\global\chapletter={D}
  \or\global\chapletter={E} \or\global\chapletter={F} \or\global\chapletter={G}
  \or\global\chapletter={H} \or\global\chapletter={I} \or\global\chapletter={J}
  \or\global\chapletter={K} \or\global\chapletter={L} \or\global\chapletter={M}
  \or\global\chapletter={N} \or\global\chapletter={O} \or\global\chapletter={P}
  \or\global\chapletter={Q} \or\global\chapletter={R} \or\global\chapletter={S}
  \or\global\chapletter={T} \or\global\chapletter={U} \or\global\chapletter={V}
  \or\global\chapletter={W} \or\global\chapletter={X} \or\global\chapletter={Y}
  \or\global\chapletter={Z}\fi}

%
\def\tempsetletter#1{\ifcase-#1 {}\or{} \or\chapletter={A} \or\chapletter={B}
 \or\chapletter={C} \or\chapletter={D} \or\chapletter={E}
  \or\chapletter={F} \or\chapletter={G} \or\chapletter={H}
   \or\chapletter={I} \or\chapletter={J} \or\chapletter={K}
    \or\chapletter={L} \or\chapletter={M} \or\chapletter={N}
     \or\chapletter={O} \or\chapletter={P} \or\chapletter={Q}
      \or\chapletter={R} \or\chapletter={S} \or\chapletter={T}
       \or\chapletter={U} \or\chapletter={V} \or\chapletter={W}
        \or\chapletter={X} \or\chapletter={Y} \or\chapletter={Z}\fi}

%
\def\chapshow#1{
    \ifnum #1>0 \relax
        #1
    \else
        {\tempsetletter{\number#1}\the\chapletter}
    \fi}

%
\def\today{\number\day\space \ifcase\month\or January\or February\or
        March\or April\or May\or June\or July\or August\or September\or
        October\or November\or December\fi, \space\number\year}

\def\timeofday{\minutes=\time    \hours=\time
        \divide \hours by 60
        \multiply \hours by 60
        \advance \minutes by -\hours
        \divide \hours by 60
        \ifnum\the\minutes>9
     		\relax\the\hours:\the\minutes
 	\else
  		\relax\the\hours:0\the\minutes
	\fi}


%
%
%
%
\def\chapnum{\bumpchapno \chapfolio}

\def\chaplabel#1{
    \ifforwardreference                             
        \write\labelfileout{                        
        \noexpand\expandafter\noexpand\def          
        \noexpand\csname CHAPLABEL#1\endcsname{\the\chapno}}
    \fi
    \global\expandafter\edef\csname CHAPLABEL#1\endcsname
    {\the\chapno}
    \ifproofmode
        \rlap{\hbox{\marginstyle #1\ }}
    \fi}

%
\def\sectnum{
    \bumpsectno
        \ifchaptersectionnumbers
            \chapfolio.
        \fi
    \the\sectno}

\def\sectlabel#1{
    \bumpsectno
    \ifforwardreference
        \immediate\write\labelfileout{
        \noexpand\expandafter\noexpand\def
        \noexpand\csname SECTLABEL#1\endcsname{\the\chapno.\the\sectno?!}}
    \fi
    \global\expandafter\edef\csname SECTLABEL#1\endcsname
    {\the\chapno.\the\sectno?!}	 			
    \ifproofmode
        \llap{\hbox{\marginstyle #1\ }}
    \fi
    \ifchaptersectionnumbers
        \chapfolio.
    \fi
    \the\sectno}

\def\sectref#1{                                  
    \ifundefined{SECTLABEL#1}                     
        ++                                        
        \ifproofmode
            \ifforwardreference
            \else
            \write16{ ***Undefined\space Section\space Reference\space #1*** }
            \fi
        \else
        \write16{ ***Undefined\space Section\space Reference\space #1*** }
        \fi
    \else
        \edef\LABxx{\getlabel{SECTLABEL#1}}
	\ifchaptersectionnumbers
            \def\LAByy{\expandafter\stripchap\LABxx}
	    \chapshow\LAByy.
	\fi
	\expandafter\stripsect\LABxx
    \fi
    \ifcrossreference
        \write\allcrossfile{Section\space #1}
    \fi}

%
%
\def\eqnum{                                    
    \global\advance\eqlabelno by 1              
    \eqno(
    \ifeqchapternumbers
        \chapfolio.
    \fi
    \ifeqsectionnumbers
        \the\sectno.
    \fi
    \the\eqlabelno)}

\def\eqlabel#1{                                
    \global\advance\eqlabelno by 1              
    \ifforwardreference                     
        \immediate\write\labelfileout{\noexpand\expandafter\noexpand\def
        \noexpand\csname EQLABEL#1\endcsname
        {\the\chapno.\the\sectno?\the\eqlabelno!}}
    \fi
    \global\expandafter\edef\csname EQLABEL#1\endcsname
    {\the\chapno.\the\sectno?\the\eqlabelno!}
    \eqno(
    \ifeqchapternumbers
        \chapfolio.
    \fi
    \ifeqsectionnumbers
        \the\sectno.
    \fi
    \the\eqlabelno)
    \ifproofmode
        \rlap{\hbox{\marginstyle #1}}		
    \fi}

\def\eqalignnum{                               
    \global\advance\eqlabelno by 1              
    &(\ifeqchapternumbers
        \chapfolio.
    \fi
    \ifeqsectionnumbers
        \the\sectno.
    \fi
    \the\eqlabelno)}

\def\eqalignlabel#1{                   	
    \global\advance\eqlabelno by 1 	        
    \ifforwardreference                     
        \immediate\write\labelfileout{\noexpand\expandafter\noexpand\def
        \noexpand\csname EQLABEL#1\endcsname
        {\the\chapno.\the\sectno?\the\eqlabelno!}}
    \fi
    \global\expandafter\edef\csname EQLABEL#1\endcsname
    {\the\chapno.\the\sectno?\the\eqlabelno!}
    &(\ifeqchapternumbers
        \chapfolio.
    \fi
    \ifeqsectionnumbers
        \the\sectno.
    \fi
    \the\eqlabelno)
    \ifproofmode
        \rlap{\hbox{\marginstyle #1}}			
    \fi}

\def\dnum{                                     
    \global\advance\eqlabelno by 1              
    \llap{(	 				
    \ifeqchapternumbers
        \chapfolio.
    \fi
    \ifeqsectionnumbers
        \the\sectno.
    \fi
    \the\eqlabelno)}}

\def\dlabel#1{                                 
    \global\advance\eqlabelno by 1              
    \ifforwardreference                         
        \immediate\write\labelfileout{\noexpand\expandafter\noexpand\def
        \noexpand\csname EQLABEL#1\endcsname
        {\the\chapno.\the\sectno?\the\eqlabelno!}}
    \fi
    \global\expandafter\edef\csname EQLABEL#1\endcsname
    {\the\chapno.\the\sectno?\the\eqlabelno!}
    \llap{(
    \ifeqchapternumbers
        \chapfolio.
    \fi
    \ifeqsectionnumbers
        \the\sectno.
    \fi
    \the\eqlabelno)}
    \ifproofmode
        \rlap{\hbox{\marginstyle #1}}		
    \fi}

\def\eqref#1{\ifparenequations(\fi
    \ifundefined{EQLABEL#1}***
        \ifproofmode
            \ifforwardreference
            \else
            \write16{
                ***Undefined\space Equation\space Reference\space #1*** }
            \fi
        \else
        \write16{ ***Undefined\space Equation\space Reference\space #1*** }
        \fi
    \else
        \edef\LABxx{\getlabel{EQLABEL#1}}
	\def\LAByy{\expandafter\stripsect\LABxx}
        \def\LABzz{\expandafter\stripchap\LABxx}
        \ifeqchapternumbers
            \chapshow{\LABzz}.
        \else
            \ifnum \number\LABzz=\chapno \relax
            \else
                \chapshow{\LABzz}.
            \fi
        \fi
	\ifeqsectionnumbers
	    \LAByy.
	\fi
        \expandafter\stripeq\LABxx
    \fi
    \ifparenequations)\fi
    \ifcrossreference
        \write\allcrossfile{Equation\space #1}
    \fi}

%
\def\fignum{                                   
    \global\advance\figureno by 1\relax         
    \iffigurechapternumbers
        \chapfolio.
    \fi
    \the\figureno}

\def\figlabel#1{				
    \global\advance\figureno by 1\relax 	
    \ifforwardreference				
        \immediate\write\labelfileout{\noexpand\expandafter\noexpand\def
        \noexpand\csname FIGLABEL#1\endcsname
        {\the\chapno.\the\sectno?\the\figureno!}}
    \fi
    \global\expandafter\edef\csname FIGLABEL#1\endcsname
    {\the\chapno.\the\sectno?\the\figureno!}
    \iffigurechapternumbers
        \chapfolio.
    \fi
    \ifproofmode
        \llap{\hbox{\marginstyle #1\ }}\relax
    \fi
    \the\figureno}

\def\figref#1{					
    \ifundefined				
        {FIGLABEL#1}!!!!			
        \ifproofmode
            \ifforwardreference
            \else
            \write16{
                ***Undefined\space Figure\space Reference\space #1*** }
            \fi
        \else
        \write16{ ***Undefined\space Figure\space Reference\space #1*** }
        \fi
    \else
        \edef\LABxx{\getlabel{FIGLABEL#1}}
        \def\LABzz{\expandafter\stripchap\LABxx}
        \iffigurechapternumbers
            \chapshow{\LABzz}.\expandafter\stripeq\LABxx
        \else \ifnum\number\LABzz=\chapno \relax
                \expandafter\stripeq\LABxx
            \else
                \chapshow{\LABzz}.\expandafter\stripeq\LABxx
            \fi
        \fi
        \ifcrossreference
            \write\allcrossfile{Figure\space #1}
        \fi
    \fi}

%
%
\def\pagelabel#1{
    \ifforwardreference
        \write\labelfileout{
        \noexpand\expandafter\noexpand\def
        \noexpand\csname PGLABEL#1\noexpand\endcsname{\the\pageno}}
    \fi
    \global\expandafter\edef\csname PGLABEL#1\endcsname{\the\pageno}}

\def\pageref#1{
    \ifundefined
        {PGLABEL#1}***
        \ifproofmode
        \else
        \write16{ ***Undefined\space Page\space Reference\space #1*** }
        \fi
    \else
        \csname PGLABEL#1\endcsname
    \fi
    \ifcrossreference
        \write\allcrossfile{Page\space #1}
    \fi}

%
\def\refnum{                                      
    \global\advance\referenceno by 1\relax         
    \the\referenceno}	                           

\def\internalreflabel#1{			
    \global\advance\referenceno by 1\relax 	
    \ifforwardreference				
        \immediate\write\labelfileout{\noexpand\expandafter\noexpand\def
        \noexpand\csname REFLABEL#1\endcsname
        {\the\chapno.\the\sectno?\the\referenceno!}}
    \fi
    \global\expandafter\edef\csname REFLABEL#1\endcsname
    {\the\chapno.\the\sectno?\the\figureno!}
    \ifproofmode
        \llap{\hbox{\marginstyle #1\hskip.5cm}}\relax
    \fi
    \the\referenceno}

\def\internalrefref#1{				
    \ifundefined				
        {REFLABEL#1}!!!!			
        \ifproofmode
            \ifforwardreference
            \else
            \write16{
                 ***Undefined\space Footnote\space Reference\space #1*** }
            \fi
        \else
        \write16{
             ***Undefined\space Footnote\space Reference\space #1*** }
        \fi
    \else
        \edef\LABxx{\getlabel{REFLABEL#1}}
        \def\LABzz{\expandafter\stripchap\LABxx}
        \expandafter\stripeq\LABxx
        \ifcrossreference
            \write\allcrossfile{Reference\space #1}
        \fi
    \fi}

%
\def\reflabel#1{\item{\internalreflabel{#1}.}}

%
\def\refref#1{\internalrefref{#1}}

\def\eq{\ifhmode Eq.~\else Equation~\fi}		
\def\eqs{\ifhmode Eqs.~\else Equations~\fi}

%
%
%
%

%
\def\getlabel#1{\csname#1\endcsname}
\def\ifundefined#1{\expandafter\ifx\csname#1\endcsname\relax}
\def\stripchap#1.#2?#3!{#1}			
\def\stripsect#1.#2?#3!{#2}			%
\def\stripeq#1.#2?#3!{#3}			
}  

\overfullrule = 0pt
\magnification = \magstep1
\baselineskip 14pt
\hsize = 6.0 truein
\vsize = 8.5 truein
\chapternumberstrue
\initialeqmacro

\line{\hfill UMTG-180}


\vglue 0.4 truein

\centerline{\bf EXACT $S$ MATRICES FOR INTEGRABLE QUANTUM SPIN CHAINS}
\bigskip

\medskip

\centerline{LUCA MEZINCESCU and RAFAEL I. NEPOMECHIE}
\centerline{Department of Physics, University of Miami}
\centerline{Coral Gables, FL 33124, USA}

\vskip 0.2 in

\bigskip

\centerline{ABSTRACT}

\vskip 0.2 in

{\leftskip .5truein \rightskip .5truein  \baselineskip 12pt
We begin with a review of the antiferromagnetic spin $1/2$ Heisenberg
chain. In particular, we show that the model has particle-like excitations
with spin $1/2$, and we compute the exact bulk $S$ matrix. We then review our
recent work which generalizes these results. We first
consider an integrable alternating spin $1/2$ - spin $1$ chain. In addition
to having excitations with spin $1/2$, this model also has excitations with
spin $0$. We compute the bulk $S$ matrix, which has some unusual features.
We then consider the open antiferromagnetic spin $1/2$ Heisenberg chain
with boundary magnetic fields. We give a direct calculation of the
boundary $S$ matrix.
(Talk presented at the conference on Statistical Mechanics and Quantum
Field Theory at USC, 16 -- 21 May 1994)
\par}

\vskip 0.2 in

\bigskip

\medskip

\noindent
{\bf \chapnum . Introduction}
\vskip 0.2truein

The investigation of integrable quantum spin chains was initiated by
Bethe${}^{\refref{bethe}}$ with the classic paper on
the closed \footnote*{One-dimensional
quantum spin chains, like strings, come in two topologies: closed
(periodic boundary conditions) and open.}
spin $1/2$ Heisenberg chain. Other examples of integrable quantum
spin chains include the open spin $1/2$
chain${}^{\refref{gaudin} - \refref{sklyanin}}$, the spin 1
chain${}^{\refref{zamolodchikov/fateev}, \refref{mnr}}$, the spin $1/2$ chain
with a spin $1$ impurity${}^{\refref{andrei/johannesson}}$, and the
alternating spin $1/2$ - spin $1$ chain${}^{\refref{devega/woynarovich}}$.

There are several motivations for studying integrable quantum spin chains.
First, these are many-body quantum mechanical models for which exact results
can be computed. Also, these models typically
have${}^{\refref{faddeev/takhtajan}, \refref{spinons}}$ a regime with
a nontrivial antiferromagnetic vacuum and novel excitations (``spinons'').
In the continuum limit, these excitations are described by $1+1$-dimensional
integrable relativistic quantum field theory. (See, e.g., Refs.
\refref{continuum}-\refref{mccoy}.) Last but
not least, such models have applications in statistical mechanics
and condensed matter physics${}^{\refref{cm}}$ and perhaps also in
string theory${}^{\refref{strings}}$.

In this talk we review our recent work on both bulk and boundary
$S$ matrices for the excitations of integrable quantum spin chains.
Such $S$ matrices provide valuable information about long-distance
physics and boundary phenomena of the models. (See, e.g., Refs.
\refref{zamolodchikov/zamolodchikov}, \refref{ghoshal/zamolodchikov}
and references therein.)

The outline of this talk is as follows. We begin with a brief review of
the closed spin $1/2$ Heisenberg chain in the antiferromagnetic regime,
with emphasis on the physical
properties which emerge from the Bethe Ansatz solution. In particular,
following Faddeev and Takhtajan${}^{\refref{faddeev/takhtajan}}$, we outline
the argument that the model has particle-like excitations with spin $1/2$.
These excitations interact, and we explain how the exact $S$ matrix can
be computed.

In the remainder of the talk, we generalize these results in two
different directions.
In Section 3 we consider the alternating spin $1/2$ - spin $1$ chain
in the antiferromagnetic regime. Following Ref. \refref{dmn2}, we show
that in addition to having excitations with spin $1/2$ (as in the Heisenberg
chain), this model also has excitations with spin 0. We compute the
$S$ matrix, which has some unusual features. In Section 4, we consider the
open antiferromagnetic spin $1/2$ Heisenberg chain with boundary magnetic
fields. Following Ref. \refref{gmn}, we give a direct calculation of the
boundary $S$ matrix. This is the first first-principles calculation of
a boundary $S$ matrix corresponding to an interacting relativistic
field theory.
Our result agrees with the boundary $S$ matrix for the
boundary sine-Gordon model with $\beta^2 \rightarrow 8\pi$ and with
``fixed'' boundary conditions${}^{\refref{ghoshal/zamolodchikov},
\refref{fendley/saleur}}$.

\vskip 0.4truein
\noindent
{\bf \chapnum . Closed Spin $1/2$ Chain}
\vskip 0.2truein

The Hamiltonian of the closed antiferromagnetic isotropic spin $1/2$
Heisenberg chain is given by
$${\cal H} = {1\over 4}\sum_{n=1}^N \left( \vec \sigma_n \cdot
\vec \sigma_{n+1} - 1 \right)  \,, \qquad\qquad
\vec \sigma_{N+1} = \vec \sigma_1 \,, \eqlabel{hamiltonian}
$$
where $\vec\sigma$ are the usual Pauli spin matrices.
We assume that the number of spins, $N$, is even. The Hamiltonian
commutes with the ``momentum'' operator $P$ (defined such that $e^{iP}$ is
the one-site shift operator), as well as with the $su(2)$ generators
$\vec S = {1\over 2}\sum_{n=1}^N \vec \sigma_n$.
The so-called Bethe Ansatz states are the simultaneous eigenstates of
${\cal H}$, $P$, $S^2$ and $S^z$ which are highest weights of $su(2)$
(i.e., with corresponding eigenvalues $S = S^z \ge 0$). These states
have been determined by both the coordinate${}^{\refref{bethe}}$ and
algebraic${}^{\refref{algebraic}}$ Bethe Ansatz methods.\footnote*{The
remaining states are obtained by acting on the Bethe Ansatz states
with the spin lowering operator $S^-$.} In the latter approach,
one constructs certain creation and destruction operators, $B(\lambda)$
and $C(\lambda)$, respectively; and the Bethe Ansatz states are given by
$$B(\lambda_1)\ B(\lambda_2) \cdots B(\lambda_M)\
\omega^+ \,, \eqnum $$
where $\omega^+$ is the ferromagnetic vacuum state with all spins up,
$$ C(\lambda)\ \omega^+ = 0 \,, \eqnum $$
and $\{ \lambda_\alpha \}$ satisfy the Bethe Ansatz (BA) equations
$$
\left({\lambda_\alpha + {i\over 2} \over
       \lambda_\alpha - {i\over 2}} \right)^N
=  \prod_{\scriptstyle{\beta=1}\atop \scriptstyle{\beta \ne \alpha}}^M
\left( {\lambda_\alpha - \lambda_\beta + i \over
        \lambda_\alpha - \lambda_\beta - i} \right) \,,
\qquad \alpha = 1, \cdots , M  \,, \qquad M \le {N\over 2} \,.
\eqlabel{BA} $$
The corresponding eigenvalues
are given by
$$\eqalignno{
E &= - {1\over 2}\sum_{\alpha=1}^M {1 \over \lambda^2_\alpha + {1\over 4}}
\,, \cr
P &= {1\over i} \sum_{\alpha=1}^M \log \left({\lambda_\alpha + {i\over 2}
\over \lambda_\alpha - {i\over 2}} \right) \,,\eqalignlabel{eigenvalues} \cr
S &= S^z = {N\over 2} - M \,.  \cr}  $$

For the ferromagnetic spin chain with Hamiltonian $-{\cal H}$, the ground
state has all spins aligned, and evidently corresponds to $M=0$. For the
antiferromagnetic spin chain with Hamiltonian $+{\cal H}$ \eqref{hamiltonian},
the identification of the ground state and the lowest-lying excited states
is as follows:

\vfill\eject

\noindent
{\it 2.1 Ground state}
\medskip

For the ground state, one can argue that $M = {N\over 2}$ and that the
roots $\{ \lambda_1 \,, \cdots \,, \lambda_M \}$ are all distinct and real.
This solution of the BA equations corresponds to a filled Fermi sea
(i.e., no holes). See Fig. 1.

For $N \rightarrow \infty$, the set of $\lambda$'s becomes dense on the
real line, and is described by the density $\sigma_{vac}(\lambda)$ which
is given by\footnote*{The Fermi points are at $\pm \infty$. Had
we introduced a bulk magnetic field, the Fermi points would be
at $\lambda = \pm \Lambda$, with $\Lambda$ finite.}
$$ \sigma_{vac}(\lambda) = {1\over 2 \cosh \pi \lambda}
+ O({1\over N^2}) \,. \eqlabel{ground} $$
(This result is obtained by solving the linear integral equation
for the root density which follows from the BA equations.)
Making in \eq\eqref{eigenvalues} the following replacement of
sums by integrals
$${1\over N} \sum_{\alpha=1}^M \left( \ \right) \rightarrow
\int_{-\infty}^\infty \left( \ \right) \sigma_{vac}(\lambda)\ d\lambda \,,
\eqnum $$
one concludes that the ground state has the following quantum numbers:
$$E = E_0 = - N \log 2 \,, \qquad P = P_0 = N \pi/2 \,, \qquad
  S = 0 \,. \eqnum $$
In particular, the ground state is a spin singlet, as one would expect for
an antiferromagnet.

\medskip
\noindent
{\it 2.2 Excitations}
\medskip

The excited states above the ground state consist of an even number of
particle-like excitations, which are now known as ``spinons''.
(Faddeev-Takhtajan called them ``kinks''.) Therefore, the lowest-lying
excited states have two spinons. One can argue that there are four such
states: the triplet ($S=1$) states, and the singlet ($S=0$) state.
(The total number of states with $\nu$ spinons is equal to $2^\nu$.)
The fact that the excited states with two spinons have $S=1$ and $S=0$
implies the important result that a spinon has spin 1/2.

The triplet state with $S=S^z=1$ is described by only real roots
$\{ \lambda_1 \,, \cdots \lambda_M \}$ (as is the ground state), but
with $M = {N\over 2} -1$. This
solution of the BA equations corresponds to a Fermi sea with two holes.
The hole rapidities are labeled $\tilde \lambda_1$, $\tilde \lambda_2$.
See Fig. 2.

The singlet state corresponds to a Fermi sea with two holes (with
rapidities $\tilde \lambda_1$ and $\tilde \lambda_2$) as well as
a ``2 - string'',
which is a set of conjugate roots $\lambda_0 \pm {i\over 2}$, with
$\lambda_0$ (the ``center'' of the 2 - string) real. For the singlet state,
the BA equations further constrain the center to be given by
$$\lambda_0 = {1\over 2}(\tilde\lambda_1 + \tilde\lambda_2) \,.
\eqlabel{center} $$
All of these features of the singlet-state solution can be seen in Fig. 3.

For both the triplet and singlet states, one can show that the density
$\sigma (\lambda)$ of real roots and holes is given by an expression
of the form
$$\sigma(\lambda) = {1\over 2 \cosh \pi \lambda} + {1\over N} r(\lambda)
+ O({1\over N^2}) \,,
\eqlabel{density} $$
where $r(\lambda)$ is a correction of order 1 to the ground
state density \eqref{ground}. Heuristically, this correction corresponds
to a ``polarization'' of the Fermi sea due to the holes and the
2-string. Explicitly,
$$\eqalignno{
r_{triplet}(\lambda) &= \sum_{\alpha=1}^2 J(\lambda - \tilde\lambda_\alpha)
\,, \cr
r_{singlet}(\lambda) &= r_{triplet}(\lambda) - a_1(\lambda - \lambda_0)
 \,,
\eqalignlabel{explicit} \cr}
$$
where
$$
J(\lambda) = {1\over 2\pi} \int_{-\infty}^\infty d\omega\
e^{-i \omega \lambda}\  {e^{- |\omega|}\over 1 + e^{-|\omega|}}
\,, \qquad
a_1(\lambda) = {i\over 2\pi} {d\over d\lambda} \log
\left({\lambda + {i\over 2} \over \lambda - {i\over 2}} \right)
\,. \eqnum $$
It follows from \eq\eqref{eigenvalues} that for both the triplet
and singlet states, the energy and momentum are given by
$$\eqalignno{
E &= E_0 + \varepsilon(\tilde\lambda_1) + \varepsilon(\tilde\lambda_2)
\,, \cr
P &= P_0 + p(\tilde\lambda_1) + p(\tilde\lambda_2)
\,, \eqalignlabel{additivity} \cr}  $$
where $E_0$ and $P_0$ are the energy and momentum of the ground state,
and
$$\eqalignno{
\varepsilon(\lambda) &= {\pi \over 2 \cosh \pi \lambda}
\,, \eqalignlabel{energy} \cr
p(\lambda) &= \tan^{-1} \sinh \pi \lambda - {\pi\over 2}
\,. \eqalignlabel{momentum} \cr} $$
{}From the additivity property displayed by \eq\eqref{additivity},
we see that the spinons indeed are particle-like
excitations, with energy $\varepsilon(\lambda)$ and
momentum $p(\lambda)$. The energy-momentum dispersion relation is
$$ \varepsilon = -{\pi\over 2}\sin p \,. \eqnum $$
Note that the spinons are gapless
($\varepsilon(\lambda) \rightarrow 0$ for
$\lambda \rightarrow \pm \infty$).

\medskip
\noindent
{\it 2.3 $S$ matrix}
\medskip

The $S$ matrix for the scattering of spinons can be calculated
exactly. Here we follow the
Korepin-Andrei-Destri${}^{\refref{korepin}, \refref{andrei/destri}}$
method. An important observation is that for a state of two spinons
with rapidities $\tilde\lambda_1$ and $\tilde\lambda_2$,
the momentum $p(\tilde\lambda_1)$ satisfies the quantization condition
$$e^{i p(\tilde\lambda_1) N}\ \check R(\tilde\lambda_1 - \tilde\lambda_2)
 = 1 \,, \eqlabel{quantcondition} $$
where $\check R$ is the 2-particle $S$ matrix
(acting in the tensor product space $C^2 \otimes C^2$), and
$N$ is the number of spins in the chain. Let $e^{i \phi}$ be an
eigenvalue of $\check R$. Then $p(\tilde\lambda_1)$ is related to
the phase shift $\phi$ by
$$p(\tilde\lambda_1) + {1\over N} \phi= {2\pi\over N} m
\,, \eqlabel{quantization} $$
where $m$ is an integer.

On the other hand, one can show that
$$p(\tilde\lambda_1) + {2\pi\over N}\int_{-\infty}^{\tilde\lambda_1}
r(\lambda)\ d\lambda + const
= {2\pi\over N} \tilde J_1 \,, \eqlabel{show} $$
where $r(\lambda)$ is the function appearing in
\eq\eqref{density}, and $\tilde J_1$ is an integer or half-odd integer.
Comparing Eqs. \eqref{quantization} and \eqref{show}, we conclude that the
phase shift $\phi$ is given by
$$\phi = 2\pi\int_{-\infty}^{\tilde\lambda_1} r(\lambda)\ d\lambda
+ const \,. \eqnum $$
Using the explicit expressions for $r(\lambda)$ for the triplet
and singlet states, we obtain (up to a rapidity-independent phase factor)
$$\eqalignno{
S_{triplet}(\lambda) &= e^{i \phi_{triplet}} =
{\Gamma(1 + {i\lambda\over 2})
\Gamma({1\over 2} - {i\lambda\over 2})\over
\Gamma(1 - {i\lambda\over 2})
\Gamma({1\over 2} + {i\lambda\over 2})} \,, \cr
S_{singlet}(\lambda) &= e^{i \phi_{singlet}} =
- {\lambda + i\over \lambda- i} S_{triplet}(\lambda) \,,
\eqalignlabel{smatrix} \cr}
$$
where $\lambda = \tilde\lambda_1 - \tilde\lambda_2$.

\vfill\eject
\noindent
{\it 2.4 Further remarks}
\medskip

It is useful to formulate the above result as a $4 \times 4$ matrix.
Since $\check R$ commutes with $su(2)$, it is a linear combination
of the identity matrix $1$ and the permutation matrix ${\cal P}$.
Moreover, $\check R$ has the eigenvalues \eqref{smatrix}. It follows
that
$$R(\lambda) \equiv {\cal P} \check R(\lambda) =
{\Gamma(1 + {i\lambda\over 2})
\Gamma({1\over 2} - {i\lambda\over 2})\over
\Gamma(1 - {i\lambda\over 2})
\Gamma({1\over 2} + {i\lambda\over 2})}
{\left( \lambda 1 - i {\cal P} \right)\over ( \lambda - i )} \,.
\eqlabel{bulk1} $$
The matrix $R$ satisfies unitarity and crossing,
as well as the Yang-Baxter equation
$$
R_{12}(\lambda - \lambda')\  R_{13}(\lambda)\ R_{23}(\lambda') =
R_{23}(\lambda')\  R_{13}(\lambda)\ R_{12}(\lambda - \lambda')
\,,  \eqlabel{yang-baxter}
$$
where $R_{12}$, $R_{13}$, and $R_{23}$ are matrices
acting in the tensor product space $C^2 \otimes C^2 \otimes C^2$, with
$R_{12} = R \otimes 1$, $R_{23} = 1 \otimes R$, etc.
(See, e.g., Refs. \refref{zamolodchikov/zamolodchikov}, \refref{reviews}
and references therein.)

We have described here only the
calculation of the 2-particle $S$ matrix. In principle one can compute
in similar fashion the multiparticle $S$ matrix, and verify that
the multiparticle $S$ matrix is factorizable into a product of
2-particle $S$ matrices.

Finally, we briefly discuss the continuum limit of this model.
The continuum quantum field theory is${}^{\refref{critical}}$
the $su(2)$ WZW model${}^{\refref{wzw}}$ of level $k=1$.
This is an $su(2)$-invariant CFT${}^{\refref{bpz}}$ with
central charge $c=1$. Indeed, the massless $S$
matrix${}^{\refref{massless}}$ of the latter model coincides
with \eq\eqref{smatrix}.

The $S$ matrix \eqref{smatrix} can also be obtained by starting
with an anisotropic spin chain with anisotropy parameter
$\eta$ and lattice spacing $a$, and then taking the continuum limit
$a \rightarrow 0$ and the isotropic limit $\eta \rightarrow 0$
while keeping a mass parameter $m^2 \propto a^{-2} \exp (- \pi^2/\eta)$
fixed. (See Ref. \refref{continuum}.) Thus, this $S$ matrix
also describes a massive $su(2)$-invariant integrable quantum field theory
(namely${}^{\refref{zamolodchikov/zamolodchikov}}$, the $su(2)$-invariant
Thirring model, or the sine-Gordon model in the limit
$\beta^2 \rightarrow 8\pi$), which in the ultraviolet limit $m \rightarrow 0$
reduces to the WZW model.

\vskip 0.4truein
\noindent
{\bf \chapnum . Closed Alternating Spin $1/2$ - Spin $1$ Chain}
\vskip 0.2truein

We now consider a system with a strictly alternating arrangement of
$2N$ spins, with spins 1/2 at even sites and spins 1 at odd sites.
That is, there are $N$ spins
${1\over 2}\vec \sigma_2 \,, {1\over 2}\vec \sigma_4 \,, \cdots \,,
{1\over 2}\vec \sigma_{2N}$
of spin 1/2 and $N$ spins $\vec s_1 \,, \vec s_3 \,,$
$\cdots \,, \vec s_{2N-1}$  of spin 1.
The $su(2)$-invariant Hamiltonian ${\cal H}$ is given
by${}^{\refref{devega/woynarovich},\refref{dmn1}}$
$$\eqalignno{
{\cal H} =  -{1 \over 18} \sum_{n=1}^N \Big\{ &
\left( 2 \vec \sigma_{2n} \cdot \vec s_{2n+1} + 1 \right)
\left( 2 \vec \sigma_{2n+2} \cdot \vec s_{2n+1} + 3 \right)
\eqalignlabel{althamiltonian} \cr
& + \left( 2 \vec \sigma_{2n} \cdot \vec s_{2n-1} + 1 \right)
\left[ \left( 1 +  \vec s_{2n-1} \cdot \vec s_{2n+1}\right)
\left( 2 \vec \sigma_{2n} \cdot \vec s_{2n+1} + 1 \right) + 2 \right]
\Big\} \,.  \cr} $$
Note that the Hamiltonian contains both nearest
and next-to-nearest neighbor interactions. We assume periodic boundary
conditions ($\vec \sigma_{2n} \equiv \vec \sigma_{2n + 2N}$ and
$\vec s_{2n+1} \equiv \vec s_{2n +1 + 2N}$) and that $N$ is even.

It is not difficult to understand the origin of this Hamiltonian.
Using $R$ matrices\footnote*{We denote by $R^{(s_1 \,, s_2)}(\lambda)$
the $su(2)$-invariant $R$ matrix acting on the tensor product space
$C^{2 s_1 + 1} \otimes C^{2 s_2 + 1}$ corresponding to spins
$s_1$ and $s_2$. See Refs. \refref{zamolodchikov/fateev},
\refref{kulish/sklyanin}.}
$R^{({1\over 2} \,, {1\over 2})}$, $R^{({1\over 2} \,, 1)}$,
$R^{(1 \,, {1\over 2})}$ and $R^{(1 \,, 1)}$ as vertex weights, one
can construct${}^{\refref{devega/woynarovich}}$ an integrable two-dimensional
classical statistical mechanical vertex model as shown in Fig. 4.
Note that both rows and columns alternate between spin $1/2$ and spin $1$,
and that the lattice is invariant under rotation by $\pi/2$.
\noindent
The logarithmic derivative of the (two-row to two-row) transfer matrix
gives the above Hamiltonian.

The Bethe Ansatz states have been determined in
Ref. \refref{devega/woynarovich}. The corresponding energy, momentum,
and spin eigenvalues are given by
$$\eqalignno{
E &= - \sum_{\alpha=1}^M \left(
{1\over 2}{1\over \lambda_\alpha^2 + {1\over 4}}
+ {1\over \lambda_\alpha^2 + 1} \right)
+ {\hbox{independent of }} \{\lambda_\alpha\} \,, \cr
P &= {1\over 2i} \sum_{\alpha=1}^M \log
\left( {\lambda_\alpha + {i\over 2} \over \lambda_\alpha - {i\over 2}}
       {\lambda_\alpha + i \over \lambda_\alpha - i} \right)
\,, \eqalignlabel{alteigenvalues} \cr
S &= S^z = {3N\over 2} - M \,, \cr} $$
where the variables $\lambda_\alpha$ satisfy the BA equations
$$
\left( {\lambda_\alpha + {i\over 2} \over \lambda_\alpha - {i\over 2}}
       {\lambda_\alpha + i \over \lambda_\alpha - i} \right)^N
= \prod_{\scriptstyle{\beta=1}\atop \scriptstyle{\beta \ne \alpha}}^M
{\lambda_\alpha - \lambda_\beta + i \over
\lambda_\alpha - \lambda_\beta - i } \,, \qquad
\alpha = 1, \cdots , M \,, \qquad M \le {3N\over 2}
\,. \eqlabel{altBA} $$
The momentum operator is defined such that $e^{i2P}$ is the
two-site shift operator, and hence the factor $1/2$ in
\eq\eqref{alteigenvalues}.

The ground state corresponds to {\it two} filled Fermi seas:
a sea of 1-strings (i.e., real roots of the BA equations, as in the ground
state of the spin $1/2$ chain) and a sea of 2-strings.
See Fig. 5.

Holes in the sea of 2-strings are excitations with spin $1/2$, just like
the excitations of the spin $1/2$ chain. However, for the alternating
spin chain, there is the additional possibility of having holes in the
sea of 1-strings. As shown in Ref. \refref{dmn2},
holes in the sea of 1-strings are excitations with spin $0$.
To the best of our knowledge, this is the first example of a magnetic
chain with spin $0$ excitations.

For both the spin $1/2$ and spin $0$ excitations, the energy
$\varepsilon(\lambda)$ is given by \eqref{energy}, and the momentum is given
$p(\lambda)/2$, where $p(\lambda)$ is given by \eqref{momentum}.
The total number of states with $\nu$ excitations is
$\sum_{m=0}^{\nu/2} 2^{2m}$, which corresponds to having an even number
of each type of excitation.

The $S$ matrix can be computed (up to rapidity-independent phase factors)
as before.
The triplet and singlet $S$ matrix elements for the scattering of two
spin $1/2$ excitations coincide with the expressions given in
\eq\eqref{smatrix}. There is no scattering between two spin $0$ excitations
(the $S$ matrix element is $S(\lambda) = 1$)
and the $S$ matrix element for the scattering of a spin $1/2$ excitation
and a spin $0$ excitation is
$$S(\lambda) = i \coth {\pi\over 2} \left( \lambda + {i\over 2}\right)
\,, \eqnum $$
where $\lambda$ is the difference of the corresponding hole rapidities.
Remarkably, the scalar-spinor scattering is nontrivial, yet the
spinor-spinor scattering is the same as for the Heisenberg chain.

An interesting open problem is to determine the continuum limit
of this model. We know that the continuum quantum field theory
must be some $su(2)$-invariant CFT
with${}^{\refref{dmn1}, \refref{aladim/martins}}$ central charge $c=2$.

We remark that for both the spin $1/2$ chain and the alternating
spin $1/2$ - spin $1$ chain, the ratio $C_H/T$ (the specific heat at
constant field divided by the temperature) has the property
$$ \lim_{T \rightarrow 0} \lim_{H \rightarrow 0} {C_H\over T} =
\lim_{H\rightarrow 0} \lim_{T \rightarrow 0} {C_H\over T}
\,. \eqlabel{commute} $$
The LHS can be evaluated by the method of
Filyov, {\it et al.}${}^{\refref{filyov}}$
while the RHS can be evaluated by the method of Johnson and
McCoy${}^{\refref{johnson/mccoy}}$.

For integrable isotropic spin $s$ chains with $s>1/2$
${}^{\refref{kulish/sklyanin},\refref{takhtajan}, \refref{babujian}}$,
the property \eqref{commute} is not satisfied. Indeed,
the LHS is proportional to $c= 3s/(s+1)$ (see Ref. \refref{critical}),
while the RHS is proportional to $c=1$ (see Ref.
\refref{izergin/korepin/reshetikhin}).
Moreover, there is a discrepancy between the results of
Takhtajan${}^{\refref{takhtajan}}$ (see also Ref. \refref{andrei/destri})
and Reshetikhin${}^{\refref{reshetikhin}}$ for the two-body $S$ matrix:
$$S_{Takhtajan} \ne S_{Reshetikhin} \,. \eqnum $$
These facts strongly suggest that there are (at least) two continuous
field theories in the $(T\,, H) = (0 \,, 0)$ limit of the spin $s$
isotropic chain. The limit $T=0 \,, H=0^+$ corresponds to a $c=1$ theory
with Takhtajan's $S$ matrix; and the limit $H=0 \,, T=0^+$ corresponds to
a $c= 3s/(s+1)$ theory with Reshetikhin's $S$ matrix.

\vskip 0.4truein
\noindent
{\bf \chapnum . Open Spin $1/2$ Chain with Boundary Magnetic Fields}
\vskip 0.2truein

We now consider the open antiferromagnetic isotropic spin $1/2$
Heisenberg chain with boundary magnetic fields.
The Hamiltonian is given by
$${\cal H} = {1\over 4}\left\{ \sum_{n=1}^N \vec \sigma_n \cdot
\vec \sigma_{n+1} + {1\over \xi_-}\sigma_1^z + {1\over \xi_+}\sigma_N^z
\right\}
\,, \eqlabel{openhamiltonian}
$$
where the (real) parameters $\xi_\pm$ correspond to boundary magnetic fields.
We assume that $\xi_\pm > 1/2$ and that $N$ is even. Since the spin chain
is open, the Hamiltonian does not commute with the shift operator.
Moreover, the boundary magnetic fields break the $su(2)$ symmetry,
and so the Hamiltonian commutes only with $S^z$.

The simultaneous eigenstates of ${\cal H}$ and $S^z$ have been determined
by both the coordinate${}^{\refref{alcaraz}}$ and
algebraic${}^{\refref{sklyanin}}$ Bethe Ansatz. In the latter approach,
one constructs (in analogy with the closed spin chain)
certain creation and destruction operators, ${\cal B}(\lambda)$
and ${\cal C}(\lambda)$, respectively; and the eigenstates are given by
$${\cal B}(\lambda_1)\ {\cal B}(\lambda_2) \cdots {\cal B}(\lambda_M)\
\omega^+ \,, \eqlabel{state} $$
where $\omega^+$ is the ferromagnetic vacuum state with all spins up,
$$ {\cal C}(\lambda)\ \omega^+ = 0 \,, \eqlabel{pseudovacuum} $$
and $\{ \lambda_\alpha \}$ satisfy the Bethe Ansatz (BA) equations
$$\eqalignno{
&\left({\lambda_\alpha + i(\xi_+ -{1\over 2}) \over
       \lambda_\alpha - i(\xi_+ -{1\over 2})} \right)
\left({\lambda_\alpha + i(\xi_- -{1\over 2}) \over
       \lambda_\alpha - i(\xi_- -{1\over 2})} \right)
\left({\lambda_\alpha + {i\over 2} \over
       \lambda_\alpha - {i\over 2}} \right)^{2N} \cr
& \qquad =
\prod_{\scriptstyle{\beta=1}\atop \scriptstyle{\beta \ne \alpha}}^M
\left( {\lambda_\alpha - \lambda_\beta + i \over
        \lambda_\alpha - \lambda_\beta - i} \right)
\left( {\lambda_\alpha + \lambda_\beta + i \over
        \lambda_\alpha + \lambda_\beta - i} \right) \,,
\qquad \alpha = 1, \cdots , M  \,.
\eqalignlabel{openBA} \cr} $$
The corresponding energy and spin eigenvalues are given by
$$\eqalignno{
E &= - {1\over 2}\sum_{\alpha=1}^M {1 \over \lambda^2_\alpha + {1\over 4}}
 + {\hbox{independent of }} \{\lambda_\alpha\} \,,
\eqalignlabel{openenergy} \cr
S^z &= {N\over 2} - M \,.
\eqalignnum \cr} $$
We require that the BA solutions correspond to independent BA states,
and therefore, we make the restriction
$${\hbox{ Re}} \left( \lambda_\alpha \right) > 0 \,.  \eqnum $$
(See, e.g., Refs. \refref{gaudin}, \refref{alcaraz}, \refref{gmn},
\refref{fendley/saleur}, \refref{destri/devega}, \refref{hamer}.)

As for the closed spin $1/2$ chain, the ground state corresponds
to a real Fermi sea, and the excitations are spinons with $S^z = \pm 1/2$
and energy $\varepsilon(\lambda)$. We assume that
the 2-particle $S$ matrix is the same as for the closed spin chain. The
problem is to compute the boundary $S$ matrix, which describes the
interaction of a spinon with the end of the spin chain. However,
it is instructive to first consider a similar but more elementary
problem.

\medskip
\noindent
{\it 4.1 Boundary $S$ matrix: free particle}
\medskip

As a warm-up exercise, we first compute the boundary $S$ matrix
for a free nonrelativistic particle of mass $m$
(with Hamiltonian $H=p^2/2m$) which is constrained to be
on the positive half-line $x \ge 0$. Usually
one demands that the wavefunction $\psi(x)$ vanish at $x=0$. This
is a sufficient, but by no means necessary, condition for the probability
current $j(x) = i \psi(x)^* {\buildrel \leftrightarrow \over \partial_x}
\psi(x)$ to vanish at $x=0$. We consider
instead the more general (mixed Dirichlet-Neumann) boundary condition
$$ c\psi(x) + {d\over dx}\psi(x) = 0 \quad\quad {\hbox{  at  }} x=0 \,,
\eqlabel{bc} $$
where $c$ is a real parameter with dimension 1/length. This boundary
condition also implies the
vanishing of the probability current at $x=0$, and is compatible with
the self-adjointness of the Hamiltonian. (This boundary condition
has been shown${}^{\refref{bibikov/tarasov}}$ to be compatible with
the integrability of the nonlinear Schr\"odinger equation on the
positive half-line.) Assuming energy eigenfunctions of the plane-wave form
$$\psi_p(x) = A e^{i p x} + B e^{-i p x}   \eqlabel{planewave} $$
(we set $\hbar = 1$), we can use the boundary condition \eqref{bc} to
eliminate $A$ in terms of $B$; and we immediately obtain
$$\psi_p(x) = B \left[ e^{-i p x} + \left( {p + ic\over p -ic} \right)
e^{i p x} \right] \,. \eqnum $$
We conclude that the boundary $S$ matrix is given by
$$K(p) = {p + ic\over p -ic} \,. \eqnum $$
We see that the boundary can give rise to a nontrivial boundary $S$ matrix.
The pole at $p=ic$ implies the existence (for $c > 0$ ) of a boundary
bound state with energy $E = -c^2/2m$.


\medskip
\noindent
{\it 4.2 Boundary $S$ matrix: open spin $1/2$ chain}
\medskip

For the open spin $1/2$ chain, the quantization condition
\eqref{quantcondition} is replaced by${}^{\refref{fendley/saleur}}$
$$e^{i 2 p(\tilde\lambda_1) N}\ R_{12}(\tilde\lambda_1 - \tilde\lambda_2)\
K_1(\tilde\lambda_1 \,, \xi_-)\
R_{21}(\tilde\lambda_1 + \tilde\lambda_2)\ K_1(\tilde\lambda_1 \,, \xi_+)
= 1 \,. \eqlabel{qq3} $$
Here $p(\lambda)$ is defined by \eqref{momentum} (i.e, the expression
for the momentum of a particle with rapidity $\lambda$ for the corresponding
system with periodic boundary conditions), and $K(\lambda, \xi)$ is
the boundary $S$ matrix (acting in the space $C^2$).
We use the same notation employed in \eq\eqref{yang-baxter}; moreover,
$$R_{21}(\lambda) \equiv {\cal P}_{12}\ R_{12}(\lambda)\ {\cal P}_{12}
\,,  \eqnum $$
where ${\cal P}$ is the permutation matrix; and $K_1$, $K_2$ denote
matrices acting in the space $C^2 \otimes C^2$, with $K_1 = K \otimes 1$,
$K_2 = 1 \otimes K$.

The $R$ matrix is given in \eq\eqref{bulk1}.
This matrix has the following form
$$R(\lambda) =  \left( \matrix{ a(\lambda) &0  &0  &0 \cr
                               0  & b(\lambda) & c(\lambda) & 0 \cr
                               0  & c(\lambda) & b(\lambda) & 0 \cr
                               0  & 0     & 0  & a(\lambda) \cr} \right)
\,, \qquad \eqlabel{bulk2}
$$
with
$$b(\lambda) = {\lambda\over \lambda - i} a(\lambda) \,, \qquad
  c(\lambda) = - {i\over \lambda - i} a(\lambda) \,, \qquad
  a(\lambda) = {\Gamma(1 + {i\lambda\over 2})
\Gamma({1\over 2} - {i\lambda\over 2})\over
\Gamma(1 - {i\lambda\over 2})\Gamma({1\over 2} + {i\lambda\over 2})}
\,.$$

The $U(1)$ symmetry of the Hamiltonian's boundary terms implies that the
boundary $S$ matrix is of the form
$$K(\lambda \,, \xi) = \left( \matrix{ \alpha(\lambda \,, \xi) &0  \cr
                                     0  & \beta(\lambda \,, \xi) \cr} \right)
\,. \eqlabel{form} $$
Our task is to explicitly determine the matrix elements
$\alpha(\lambda \,, \xi)$ and $\beta(\lambda \,, \xi)$,
which are the boundary scattering amplitudes for excitations
with $S^z = +1/2$ and $S^z = -1/2$, respectively.
We proceed by examining the two-particle excited states, which
we classify by their $S^z$ eigenvalue. As for the closed spin $1/2$
chain, there are four such states ($S^z =1$, $S^z =-1$, and two
states with $S^z = 0$). Since we need to determine
only two matrix elements, the system of four equations provided
by the quantization condition \eqref{qq3} is overdetermined.
The structure \eqref{bulk2} of the $R$ matrix suggests that there
will be two simple relations corresponding to the diagonal elements
of the $R$ matrix. These relations will enable us to determine
the matrix elements $\alpha(\lambda \,, \xi)$ and
$\beta(\lambda \,, \xi)$. The other two relations should lead to
identities.

\medskip
\noindent
$S^z = 1$ {\it state}
\medskip

For the $S^z = 1$ state, the quantization condition \eqref{qq3}
implies
$$2 p(\tilde\lambda_1) + {1\over N} \Phi^{(1)} = {2 \pi\over N} m  \,,
\eqlabel{q4} $$
with
$$e^{i\Phi^{(1)}} = a(\tilde\lambda_1 - \tilde\lambda_2)\
\alpha(\tilde\lambda_1 \,, \xi_-)\ a(\tilde\lambda_1 + \tilde\lambda_2)\
\alpha(\tilde\lambda_1 \,, \xi_+)  \,. \eqlabel{qq4} $$

As for the closed spin chain, the $S^z = 1$ state is the Bethe
Ansatz state consisting of two holes in the (real) Fermi sea.
Using the BA equations, we can compute${}^{\refref{gmn}}$
the function $r(\lambda)$,
which is the sum of $1/N$ contributions to the
density $\sigma(\lambda)$ for this state.\footnote*{In contrast to the
closed-chain result \eqref{ground}, the ground-state density
$\sigma_{vac}(\lambda)$ for the open spin chain has corrections of
order $1/N$, and these corrections contribute to $r(\lambda)$.}
For the open spin chain, the identity \eqref{show}
is replaced by
$$ 2 p(\tilde\lambda_1) + {2\pi\over N}\int_0^{\tilde\lambda_1}
r(\lambda)\ d\lambda + const
= {2\pi\over N} \tilde J_1 \,. \eqlabel{openshow} $$
It follows that
$$\Phi^{(1)} = 2\pi\int_0^{\tilde\lambda_1} r(\lambda)\ d\lambda
+ const \,.  \eqnum $$
Using the explicit expressions for $r(\lambda)$ and $a(\lambda)$,
we obtain the following result for $\alpha(\lambda \,, \xi)$
(up to a rapidity-independent phase factor):
$$\alpha(\lambda \,, \xi) =
{\Gamma \left({-i\lambda\over 2} + {1\over 4}\right) \over
  \Gamma \left({i\lambda\over 2} + {1\over 4}\right)}
{\Gamma \left({i\lambda\over 2} + 1\right) \over
\Gamma \left({-i\lambda\over 2} + 1\right)}
{\Gamma \left({-i\lambda\over 2} + {1\over 4}(2\xi -1)\right)\over
  \Gamma \left({i\lambda\over 2} + {1\over 4}(2\xi -1)\right)}
{\Gamma \left({i\lambda\over 2} + {1\over 4}(2\xi +1)\right)\over
\Gamma \left({-i\lambda\over 2} + {1\over 4}(2\xi +1)\right)} \,.
\eqlabel{result1} $$

\medskip
\noindent
$S^z = -1$ {\it state}
\medskip

To determine the remaining element
$\beta(\lambda \,, \xi)$ of the boundary $S$ matrix, we consider the
$S^z = -1$ state. The quantization condition \eqref{qq3} implies
$$2 p(\tilde\lambda_1) + {1\over N} \Phi^{(-1)} = {2 \pi\over N} m
\,, \eqnum $$
with
$$e^{i\Phi^{(-1)}} = a(\tilde\lambda_1 - \tilde\lambda_2)\
\beta(\tilde\lambda_1 \,, \xi_-)\ a(\tilde\lambda_1 + \tilde\lambda_2)\
\beta(\tilde\lambda_1 \,, \xi_+) \,. \eqnum $$

The $S^z = -1$ state is most easily described within the BA approach
by changing the pseudovacuum. Hence, instead of working with the states
\eqref{state}, we work now with
$${\cal C}(\lambda_1)\ {\cal C}(\lambda_2) \cdots {\cal C}(\lambda_M)\
\omega^- \,, \eqlabel{newstates} $$
where $\omega^-$ is the ferromagnetic vacuum state with all spins down,
$$ {\cal B}(\lambda)\ \omega^- = 0 \,. \eqlabel{pseudovacuum} $$
Sklyanin has shown${}^{\refref{sklyanin}}$ that
$\{ \lambda_\alpha \}$ in \eq\eqref{newstates}
satisfy the same BA equations \eqref{openBA}
as before, except for the replacement of $\xi_\pm$ by $-\xi_\pm$.
The energy eigenvalues are given by the same expression
\eqref{openenergy}, and the $S^z$ eigenvalues are now given by
$$S^z = M - {N\over 2} \,. \eqnum  $$

The $S^z = -1$ state now corresponds to the Bethe Ansatz state
consisting of two holes in the Fermi sea. The calculation
of the function $r(\lambda)$ is exactly the same as for the
$S^z=1$ state, except that we must track the change
$\xi_\pm \rightarrow -\xi_\pm$. We find that $\beta(\lambda\,, \xi)$
is given by
$$\beta(\lambda\,, \xi) = -{\lambda +i(\xi -{1\over 2})\over
\lambda -i(\xi -{1\over 2})} \alpha(\lambda \,, \xi) \,, \eqlabel{result2} $$
where $\alpha(\lambda\,, \xi)$ is given by \eq\eqref{result1}.
This completes the derivation of the boundary $S$ matrix.

\medskip
\noindent
$S^z = 0$ {\it states}
\medskip

We have already succeeded to determine the boundary $S$ matrix.
Nevertheless, a good check on this result and on the general formalism
is provided by analyzing the $S^z = 0$ states, of which there are two.
In particular, we consider the $S^z = 0$ state consisting of two holes
in the Fermi sea, and also one 2-string.
For $\xi_\pm \rightarrow \infty$, this is the spin-singlet $(S = S^z = 0)$
state shown in Fig. 6. \footnote*{The other $S^z = 0$ state is the
one which for $\xi_\pm \rightarrow \infty$ is one of the spin triplet
$(S = 1)$ states. For $\xi_\pm \ne \infty$, it is not clear how to
identify this state in terms of the Bethe Ansatz solution, and
we do not consider it further.}
\noindent
The position $\lambda_0$ of the center of the 2-string is {\it not}
given by the simple expression \eqref{center}. For example, for the
special case $\xi_\pm = \infty$, the center position is
$$ \lambda_0 = {\sqrt{ {1\over 4} + {1\over 2}\left[
(\tilde\lambda_1)^2 + (\tilde\lambda_2)^2 \right] }} \,.
\eqlabel{opencenter} $$
The general case $\xi_\pm \ne \infty$ is discussed in Ref. \refref{gmn}.

For the $S^z = 0$ states, the quantization condition \eqref{qq3}
leads to a $2 \times 2$ matrix equation. The two eigenvalues of this matrix
are pure phases. Since the matrix elements of $R(\lambda)$ and
$K(\lambda \,, \xi_\pm)$ are known, these eigenvalues can be computed
explicitly. Let $\exp i\Phi^{(0)}$ be the eigenvalue which
for $\xi_\pm \rightarrow \infty$ corresponds to the spin-singlet
$(S = S^z = 0)$ state. The quantization condition implies
$$e^{i 2 p(\tilde\lambda_1) N} e^{i\Phi^{(0)}} = 1 \,.
\eqlabel{totalphasesinglet} $$

{}From \eq\eqref{openshow} and the corresponding function $r(\lambda)$,
we obtain the consistency condition
$$ e^{i\left( \Phi^{(0)} - \Phi^{(1)} \right)}
= e_1 (\tilde\lambda_1 - \lambda_0)\
  e_1 (\tilde\lambda_1 + \lambda_0)
\,, \eqlabel{identity2}   $$
where
$$e_1(\lambda) = {\lambda + {i\over 2} \over \lambda - {i\over 2}}\,,
\eqnum $$
$\Phi^{(1)}$ is defined in \eqref{qq4}, the hole rapidities
$\tilde\lambda_1$ and $\tilde\lambda_2$ are arbitrary, and
$\lambda_0$ is the corresponding rapidity of the center
of the 2-string.

For the case $\xi_\pm = \infty$, this relation is satisfied
by virtue of the algebraic identity
$$
e_1 \left( {1\over 2}(\tilde\lambda_1 - \tilde\lambda_2) \right)\
e_1 \left( {1\over 2}(\tilde\lambda_1 + \tilde\lambda_2) \right) =
e_1 (\tilde\lambda_1 - \lambda_0)\
e_1 (\tilde\lambda_1 + \lambda_0)
 \,. \eqlabel{identity}  $$
which is true for arbitrary values of $\tilde\lambda_1$
and $\tilde\lambda_2$, where $\lambda_0$ is given by
\eqref{opencenter}. We have explicitly verified the formula
\eqref{identity2} also for the case $\xi_- = \infty$, $\xi_+ \ne \infty$,
and presumably it is true in general. This equality provides a nontrivial
consistency check of the bulk and boundary $S$ matrices and of the general
formalism.

\medskip
\noindent
{\it 4.3 Further remarks}
\medskip

The boundary $S$ matrix $K(\lambda \,, \xi)$ given by Eqs.
\eqref{form}, \eqref{result1}, \eqref{result2}
satisfies boundary unitarity and boundary cross-unitarity
${}^{\refref{ghoshal/zamolodchikov}}$, as well as the boundary
Yang-Baxter equation${}^{\refref{sklyanin}, \refref{ghoshal/zamolodchikov},
\refref{cherednik}, \refref{nonsymmetric}}$
$$ R_{12}(\lambda - \lambda') K_1(\lambda \,, \xi)
R_{21}(\lambda + \lambda') K_2(\lambda' \,, \xi)
= K_2(\lambda'\,, \xi) R_{12}(\lambda + \lambda') K_1(\lambda\,, \xi)
R_{21}(\lambda - \lambda') \,.
\eqlabel{reflection} $$

Since the bulk $S$ matrix coincides with that of the
sine-Gordon model with $\beta^2 \rightarrow 8\pi$,
we expect that the boundary $S$ matrix $K(\lambda \,, \xi)$
should coincide with the boundary $S$ matrix
of Ghoshal and Zamolodchikov${}^{\refref{ghoshal/zamolodchikov}}$
for the boundary sine-Gordon model with $\beta^2 \rightarrow 8\pi$ and with
``fixed'' boundary conditions. (For ``fixed'' boundary conditions,
the field theory and hence the boundary $S$ matrix are $U(1)$ invariant.)
We have verified that the two boundary $S$ matrices indeed coincide,
up to a rapidity-independent scalar factor, and with some redefinitions
of variables. The bootstrap result of Ghoshal and Zamolodchikov for the
boundary sine-Gordon model with ``fixed'' boundary conditions has been
verified using the physical Bethe Ansatz approach by Fendley and
Saleur${}^{\refref{fendley/saleur}}$. Very recently, the boundary
$S$ matrix for the anisotropic spin $1/2$ chain has been calculated
by Jimbo, {\it et al.} ${}^{\refref{kyoto}}$ using the vertex operator
approach. In the isotropic limit, their result coincides with ours.

We have seen that the analysis of the $S^z=0$ states for the open spin chain
differs significantly from that of the closed spin chain. Indeed, for the open
chain, the position of the center of the 2-string is a complicated
function of the hole rapidities $\tilde\lambda_1$ and $\tilde\lambda_2$
(as well as the boundary parameters $\xi_\pm$); while for the closed chain,
the center of the string is located midway between the two holes. Naively,
one might worry that this leads to a breakdown of factorization. However,
we have seen that factorization is maintained by virtue of certain nontrivial
identities. We expect that a similar situation holds for the closed spin
chain with four or more holes.

\vskip 0.4truein
\noindent
{\bf \chapnum . Acknowledgements}
\vskip 0.2truein

The work described here was done in collaboration with H. de Vega and
M. Grisaru. This work was supported in part by the National Science
Foundation under Grant PHY-92 09978.

\vskip 0.4truein
\noindent
{\bf \chapnum . References}
\vskip 0.2truein

\reflabel{bethe}
H. Bethe, Z. Phys. {\it 71} (1931) 205.

\reflabel{gaudin}
M. Gaudin, Phys. Rev. {\it A4} (1971) 386;
{\it La fonction d'onde de Bethe} (Masson, 1983).

\reflabel{alcaraz}
F.C. Alcaraz, M.N. Barber, M.T. Batchelor, R.J. Baxter and G.R.W. Quispel,
J. Phys. {\it A20} (1987) 6397.

\reflabel{sklyanin}
E.K. Sklyanin, J. Phys. {\it A21} (1988) 2375.

\reflabel{zamolodchikov/fateev}
A.B. Zamolodchikov and V.A. Fateev, Sov. J. Nucl. Phys. {\it 32} (1980) 298.

\reflabel{mnr}
L. Mezincescu, R.I. Nepomechie and V. Rittenberg, Phys. Lett. {\it A147}
(1990) 70.

\reflabel{andrei/johannesson}
N. Andrei and H. Johannesson, Phys. Lett. {\it 100A} (1984) 108.

\reflabel{devega/woynarovich}
H.J. de Vega and F. Woynarovich, J. Phys. {\it A25} (1992) 4499.

\reflabel{faddeev/takhtajan}
L.D. Faddeev and L.A. Takhtajan, J. Sov. Math. {\it 24} (1984) 241.

\reflabel{spinons}
See papers by D. Haldane, B. McCoy and K. Schoutens in these Proceedings.

\reflabel{continuum}
B.M. McCoy and T.T. Wu, Phys. Lett. {\it 87B} (1979) 50.

\reflabel{affleck}
I. Affleck, in {\it Fields, Strings, and Critical Phenomena}
(1988 Les Houches lectures), ed. by E. Brezin and J. Zinn-Justin
(Elsevier, 1990) p. 563.

\reflabel{massless}
A.B. Zamolodchikov and Al.B. Zamolodchikov, Nucl. Phys. {\it B379}
(1992) 602;
P. Fendley, H. Saleur, and Al. B. Zamolodchikov, Int. J. Mod. Phys.
{\it A8} (1993) 5751.

\reflabel{mccoy}
B. McCoy, hep-th/9403084.

\reflabel{cm}
See, e.g., R.J. Baxter, {\it Exactly Solved Models in Statistical
Mechanics} (Academic Press, 1982);
A.M. Tsvelick and P.B. Wiegmann, Adv. in Phys. {\it 32} (1983) 453;
N. Andrei, K. Furuya and J.H. Lowenstein, Rev. Mod. Phys. {\it 55}
(1983) 331;
J.C. Bonner, in {\it Magneto-Structural Correlations in
Exchange Coupled Systems}, ed. by R.D. Willett, D. Gatteschi and
O. Kahn (Reidel, 1985) p. 157;
N. Andrei, hep-cm/9408101;
M.T. Batchelor and C.M. Yung, hep-th/9410042; cond-mat/9410082;
papers by I. Affleck, P. Fendley, M. Fisher and A. Ludwig in these
Proceedings.

\reflabel{strings}
See, e.g., C.B. Thorn, Phys. Lett. {\it 70B} (1977) 85;
R. Giles, L.D. McLerran, and C.B.Thorn, Phys. Rev. {\it D17} (1978) 2058;
I. Klebanov and L. Susskind, Nucl. Phys. {\it B309} (1988) 175;
S. Dalley, Phys. Lett. {\it B334} (1994) 61.

\reflabel{zamolodchikov/zamolodchikov}
A.B. Zamolodchikov and Al.B. Zamolodchikov, Ann. Phys. {\it 120} (1979) 253;
A.B. Zamolodchikov, Sov. Sci. Rev. {\it A2} (1980) 1.

\reflabel{ghoshal/zamolodchikov}
S. Ghoshal and A. B. Zamolodchikov, Int. J. Mod. Phys. {\it A9} (1994)
3841; {\it A9} (1994) 4353.

\reflabel{dmn2}
H.J. de Vega, L. Mezincescu and R.I. Nepomechie, Int. J. Mod.
Phys. {\it B8} (1994) 3473.

\reflabel{gmn}
M. Grisaru, L. Mezincescu and R.I. Nepomechie, J. Phys. {\it A},
in press.

\reflabel{fendley/saleur}
P. Fendley and H. Saleur, Nucl. Phys. {\it B428} (1994) 681.

\reflabel{algebraic}
P.P. Kulish and E.K. Sklyanin, Phys. Lett. {\it 70A} (1979) 461;
L.D. Faddeev and L.A. Takhtajan, Russ. Math Surv. {\it 34} (1979) 11.
For a recent review, see
V.E. Korepin, G. Izergin and N.M. Bogoliubov, {\it Quantum Inverse
Scattering Method, Correlation Functions and Algebraic Bethe Ansatz}
(Cambridge, 1993).

\reflabel{korepin}
V.E. Korepin, Theor. Math. Phys. {\it 76} (1980) 165.

\reflabel{andrei/destri}
N. Andrei and C. Destri, Nucl. Phys. {\it B231} (1984) 445.

\reflabel{reviews}
P.P. Kulish and E.K. Sklyanin, J. Sov. Math. {\it 19} (1982) 1596;
M. Jimbo, Int. J. Mod. Phys. {\it A4} (1989) 3759.

\reflabel{critical}
H.W.J. Bl\"ote, J.L. Cardy and M.P. Nightingale, Phys. Rev. Lett. {\it 56}
(1986) 742; I. Affleck, Phys. Rev. Lett. {\it 56} (1986) 746.

\reflabel{wzw}
E. Witten, Commun. Math. Phys. {\it 92} (1984) 455; S.P. Novikov, Usp. Mat.
Nauk {\it 37} (1982) 3.

\reflabel{bpz}
A.A. Belavin, A.M. Polyakov and A.B. Zamolodchikov, Nucl. Phys. {\it B241}
(1984) 333.

\reflabel{dmn1}
H.J. de Vega, L. Mezincescu and R.I. Nepomechie, Phys. Rev. {\it B49}
(1994) 13223.

\reflabel{kulish/sklyanin}
P.P. Kulish and E.K. Sklyanin, {\it Lecture Notes in Physics} {\it 151}
(Springer, 1982) 61;
P.P. Kulish, N.Yu. Reshetikhin and E.K. Sklyanin, Lett. Math. Phys. {\it 5}
(1981) 393;
M. Jimbo, Lett. Math. Phys. {\it 10} (1985) 63.

\reflabel{aladim/martins}
S.R. Aladim and M.J. Martins, J. Phys. {\it A26} (1993) L529.

\reflabel{filyov}
V.M. Filyov, A.M. Tsvelik and P.B. Wiegmann, Phys. Lett. {\it 81A}
(1981) 175.

\reflabel{johnson/mccoy}
J.D. Johnson and B.M. McCoy, Phys. Rev. {\it A6} (1972) 1613.
For a recent review, see L. Mezincescu and R.I. Nepomechie, in
{\it Quantum Groups, Integrable Models and Statistical Systems},
ed. by J. Le Tourneux and L. Vinet (World Scientific, 1993) p. 168.

\reflabel{takhtajan}
L.A. Takhtajan, Phys. Lett. {\it 87A} (1982) 479.

\reflabel{babujian}
H.M. Babujian, Nucl. Phys. {\it B215} (1983) 317.

\reflabel{izergin/korepin/reshetikhin}
A.G. Izergin, V.E. Korepin, and N.Yu. Reshetikhin, J. Phys. {\it A22}
(1989) 2615; L. Mezincescu and R.I. Nepomechie, unpublished.

\reflabel{reshetikhin}
N. Reshetikhin, J. Phys. {\it A24} (1991) 3299.

\reflabel{destri/devega}
C. Destri and H.J. de Vega, Nucl. Phys. {\it B374} (1992) 692;
{\it B385} (1992) 361.

\reflabel{hamer}
C.J. Hamer, G.R.W. Quispel and M.T. Batchelor, J. Phys. {\it A20} (1987) 5677.
C.J. Hamer and M.T. Batchelor, J. Phys. {\it A21} (1988) L173;
A.L. Owczarek and R.J. Baxter, J. Phys. {\it A22} (1989) 1141;
M.T. Batchelor and C.J. Hamer, J. Phys. {\it A23} (1990) 761.

\reflabel{bibikov/tarasov}
P.N. Bibikov and V.O. Tarasov, Theor. Math. Phys. {\it 79} (1989) 570.

\reflabel{cherednik}
I.V. Cherednik, Theor. Math. Phys. {\it 61} (1984) 977.

\reflabel{nonsymmetric}
L. Mezincescu and R.I. Nepomechie, J. Phys. {\it A24} (1991) L17;
Int. J. Mod. Phys. {\it A6} (1991) 5231; {\it A7} (1992) 5657.

\reflabel{kyoto}
M. Jimbo, R. Kedem, T. Kojima, H. Konno and T. Miwa, hep-th/9411112.

\bigskip

\vskip 0.4truein
\noindent
{\bf Figure Captions}
\vskip 0.2truein

\item{}Fig. 1: Ground state of spin $1/2$ chain, with $N=30$.
Diamonds denote real roots of the Bethe Ansatz equations.

\medskip

\item{}Fig. 2: $S=S^z=1$ excited state. Open circles denote holes.
(The scale here differs from the one in Fig. 1.)

\medskip

\item{}Fig. 3: $S=S^z=0$ excited state. The 2-string (denoted by X's)
has its center at $(\tilde\lambda_1 + \tilde\lambda_2)/2$.

\medskip

\item{}Fig. 4: Two-dimensional vertex model. Solid and dashed
lines correspond to spin $1/2$ and spin $1$, respectively. The vertex
formed by lines corresponding to spins $s_1$ and $s_2$ has
weight $R^{(s_1 \,, s_2)}(\lambda)$. (Periodic boundary conditions should be
imposed in both horizontal and vertical directions.)

\medskip

\item{}Fig. 5: Ground state of alternating spin $1/2$ - spin $1$ chain.
Diamonds denote real roots (1-strings) and X's denote complex
roots (2-strings) of Bethe Ansatz equations.

\medskip

\item{}Fig. 6: $S=S^z=0$ excited state. The center of the 2-string is
given by \eq\eqref{opencenter}.

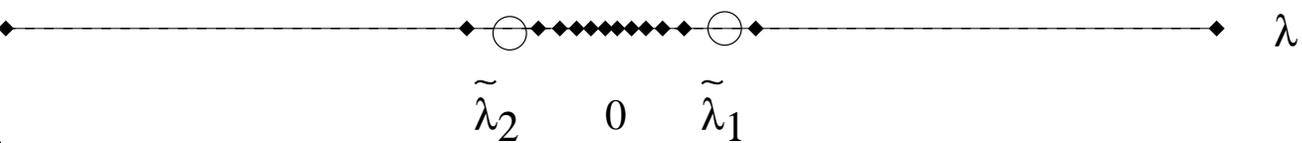
\end